\documentclass[12pt,preprint]{aastex}

\slugcomment{Submitted to MNRAS}
\shorttitle{H$_2$ emission of primordial objects}
\shortauthors{H. Kamaya \& J. Silk}

\begin{document}

\centerline{\bf  ON THE  POSSIBILITY OF OBSERVING H$_2$ EMISSION}
\centerline{\bf  FROM PRIMORDIAL MOLECULAR CLOUD KERNELS}
\bigskip
\centerline{Hideyuki Kamaya$^{1,2}$ and Joseph Silk$^{1}$}
\centerline{\it 1:Astrophysics, Department of Physics,
University of Oxford, Oxford OX1 3RH, UK}
\centerline{\it 2:Department of Astronomy, School of Science,  
Kyoto University, Kyoto, 606-8502, Japan}
\centerline{\it kamaya@astro.ox.ac.uk, silk@astro.ox.ac.uk}

\newpage

\section*{ABSTRACT}

We study  the   prospects for
observing H$_2$ emission during  the  assembly of
primordial molecular cloud kernels.  The primordial molecular cloud
cores, which  resemble  those at the present epoch,
can emerge around  $1+z=20$ according to  recent numerical simulations.
The kernels form inside the cores, and the first stars will appear
inside the kernels.  A kernel typically contracts to form 
one of the first generation   stars  with 
an accretion rate that is as large 
 as $\sim 0.01 M_\odot$ year$^{-1}$.  This occurs
due to the primordial abundances that result in a kernel temperature of order 
1000K,  and the collapsing kernel emits H$_2$ line radiation
at a rate  $\sim 10^{35}$ erg sec$^{-1}$.
Principally  $J=5-3$ (v=0) rotational emission of
H$_2$ is expected.  At  redshift  $1+z=20$, the expected flux is
$\sim 0.01~\mu$Jy for a single kernel.  While an individual
 object is not
observable  by any facilities available in 
the near future, the expected 
 assembly of primordial star clusters on  sub-galactic scales
can result in fluxes at the 
 sub-mJy level. This is marginally observable with 
 ASTRO-F.  We also examine the rotational $J=2-0$
(v=0) and vibrational $\delta v = 1$ emission lines.  The former may 
possibly  be detectable with   ALMA.

\bigskip


{\bf Key words}{: cosmology: observations --- galaxies: formation 
       --- infrared: galaxies --- ISM: molecules  
       --- submillimeter }

\newpage

\section{INTRODUCTION}

Observation
of the first generation of stars presents  one of the most exciting challenges in  
astrophysics and cosmology. Hydrogen molecules 
(H$_2$s) play an  important role as a cooling agent of gravitationally
contracting primordial gas (Saslaw \& Ziploy 1967) in the process of primordial star formation.  It has previously been argued 
 that H$_2$ is an  effective coolant for the formation of a
 globular
cluster (Peebles \& Dicke 1969), if such objects  precede galaxy formation. The contraction of 
primordial gas was also studied in the pioneering work of Matsuda, Sato,
\& Takeda (1969).  The observational feedback from  such  Population
III objects was examined in Carr, Bond, \& Arnett (1984).  In
the context of the CDM (Cold Dark Matter) scenario for cosmic structure
formation, Couchman \& Rees (1986) suggested that the feedback from the
first structure is not negligible for e.g. the reionization of the
Universe and  the Jeans mass at this epoch
(e.g. Haiman, Thoul, \& Loeb 1996; Gnedin \& Ostriker 1997; 
Ferrara 1998).  Recent progress on the role of primordial
stars formation in 
structure formation is reviewed in Nishi et al. (1998).

More recently, 
the first structures in the Universe have been studied by means
of very high  resolution numerical simulations  (Abel, Bryan, \&
Norman 2000).
The numerical resolution is
sufficient to study the formation of the first generation of molecular
clouds. According to their results, a molecular cloud
emerges with a mass of $\sim 10^5$ solar masses as a result of
the merging of small clumps which trace the initial perturbations for 
cosmic structure formation. Due to the cooling of H$_2$, a small and
cold prestellar object appears inside the primordial molecular
cloud. It  resembles  the core of molecular clouds at the present
epoch.  Their numerical results are consistent with other
numerical simulations by Bromm, Coppi, \& Larson (1999), whose results are   consistent with those  of Abel et
al. (Bromm, Coppi, \& Larson 2001).  All the simulations predict that
the primordial molecular clouds and their cores appear at the epoch of
$1+z \sim 20$ ($z$ is redshift).

According to these results, the  first
generation of  young stellar objects has a  mass of $\sim$ 200
solar mass, density of $\sim 10^5$ cm$^{-3}$, and temperature of
$\sim$ 200 K. We stress that these are  cloud cores, and not stars.
The  resolution of
the existing numerical simulations is  inadequate to 
extend the dynamic range to actual star formation.
Resort must be made to analytical arguments.

Inside the cores, a very dense structure appears. We shall
call this a {\it kernel} for clarity of  presentation.  Its density 
increases to a value as high as  $\sim 10^8$ cm$^{-3}$ where the three-body
reaction for H$_2$ formation occurs.  Further evolution is by 
fragmentation (Palla, Salpeter, \& Stahler 1983) and/or collapse
(Omukai \& Nishi 1998). In the current paper, we consider the case
that the kernel collapses: the alternative  case of fragmentation will
be 
discussed in another paper. Recent theoretical work has suggested
that  fragmentation occurs if the metallicity is above $10^{-3.5}
\times Z_{\odot}$ where $Z_{\odot}$ is the solar metallicity
(Bromm et al. 2001).

Omukai \& Nishi (1998) show that the kernel
collapses with a very large accretion rate.  It is about 0.01 $M_\odot$
year$^{-1}$.  This accretion rate is so high  due to the cooling of
H$_2$. The  H$_2$ rotational and vibrational level
transitions have wavelengths in the the rest frame
infrared band.  
The recent numerical simulations suggest that these are emitted at $1+z
\sim 20$. The redshifted wavelength  is detectable in the
infrared -- sub-mm bands. Of course, the predicted  luminosity
of individual kernels would not be
expected to be extraordinary huge, although  if the
kernels  are clustered,
the assembly of kernels may be detected by future observational
facilities.  Hence, in this paper, we will consider  
observationally feasible conditions for the assembly of primordial kernels.
Future telescope projects  that are relevant are
ALMA for the sub-mm range and ASTRO-F for the infrared range.

In \S2, we estimate the expected luminosity from a single
contracting kernel.  In \S3, we confirmed that the luminosity from
the kernel is primarily due to the emission of  H$_2$,
and  we examine which
lines are dominant.  In \S4, the observational feasibility is
discussed. 
\section{LUMINOSITY OF PRIMORDIAL MOLECULAR CLOUD KERNEL}

According to the recent numerical simulations, molecular cloud cores
appear prior to the formation of population III objects. Cores contain 
a dense and cool kernel, which
has nearly the
same physical properties  as the initial conditions chosen
by  Omukai \& Nishi (1998).
 Omukai \& Nishi find that the kernel collapses in a freely
falling manner and that the first prestellar object grows with 
a very high
accretion rate. This is possible because of the cooling by line
emission of H$_2$.  As a result, they find that the accretion rate is
$\sim $ 10$^{-2}$ $M_\odot$ year$^{-1}$.

We are interested  here in the  possibility
observing primordial  prestellar objects.  We express  
the accretion rate in  terms of luminosity. It is convenient for us to
estimate the rate of energy release owing to the high primordial accretion
rate. To determine this, we must examine the gravitational energy where the
accretion occurs.  To examine the gravitational potential energy, we
need to determine the mass distribution around the  centre of the kernel, where
the first star emerges. Also, we assume a spherical configuration for the
mass distribution.  According to Omukai \& Nishi, a high
accretion rate is realized if a similarity collapse occurs with the
adiabatic heat ratio of 1.1 (e.g. Suto \& Silk 1988).  
The density distribution is described as 
$$ \frac{\partial {\rm ln} \rho (r)}{\partial {\rm ln} r} 
= \frac{-2}{2-\gamma} . 
\eqno(1)$$
Here, $r$ is the radial distance  from the centre, $\rho (r)$ is 
the mass density
of atomic H, H$_2$ and He, and $\gamma $ is the specific heat ratio.  We
set the  mass-density distribution of a protostellar-kernel with
$\gamma = 1.1$ as $\rho_{\rm k}(r) = \rho_{\rm k0}(r/r_0)^{-2.2}$
where $\rho_{\rm k0}$ is 2.0$\times 10^{-16}$ g cm$^{-3}$ and 
$r_0$ is 0.01 pc.  
These values are appropriate for fitting the protostellar kernel of Omukai \& Nishi.

Once the density distribution of the kernel is given, we can estimate
the gravitational potential at a radius  $r$: $\Phi (r) = GM(r)/r$
where $G$ is the gravitational constant and $M(r)$ is the mass inside the
radius $r$.  Since the density is proportional to $r^{-2.2}$, we
find
$\Phi (r) = GM(r)/r = 4 \pi G \rho_{\rm k0} r_0^{2.2} / 0.8 r^{0.2}$. 
We are interested in the mass within $r_0$, then,
$\Phi (r_0) = 2.0 \times 10^{11} $ erg mass$^{-1}$ is obtained.
Hence, as found from the dimension of ''erg mass$^{-1}$'' for the
potential, we obtain the energy release rate corresponding to the
accretion rate as being
$$ L_{\rm acc} =  \dot{M} \times \Phi (r_0) =
1.3 \times 10^{35}  {\rm erg~s^{-1}} 
\times \frac{\dot{M}}{0.01 M_\odot {\rm year^{-1}} }
\eqno(2)$$
As found in the next section, this release of energy is explained by
the cooling of H$_2$ radiation.

\section{H$_2$ EMISSION  LINES}

The large accretion rate of the first stellar object arises because of
the cooling of H$_2$.  Here, we also re-formulate the energy loss
owing to line emission of H$_2$.  To estimate the cooling by 
H$_2$, we should know the number fraction of H$_2$, $f_2(r)$, and the
temperature of gas, $T(r)$, in the kernel.  Both should have 
some spatial
dependence.  We use the  fitting formula for Omukai (2000) which
is adapted from Omukai \& Nishi (1998).  
For $f_2(r)$ (solid line in Fig.1);
$$ f_2(r) = 0.0001 + 0.495 \times 
\frac{ {\rm exp}\left( \frac{n(r)}{10^{11.0} {\rm cm^{-3}}} \right)
      -{\rm exp}\left( \frac{n(r)}{10^{11.0} {\rm cm^{-3}}} \right)}
     { {\rm exp}\left( \frac{n(r)}{10^{11.0} {\rm cm^{-3}}} \right)
      +{\rm exp}\left( \frac{n(r)}{10^{11.0} {\rm cm^{-3}}} \right)}
 \eqno(3) $$
where $n(r) = 10.0^8 {\rm cm^{-3}} (r/0.01~{\rm pc})^{-2.2}$
(the maximum of $f_2(r)$ is set to be 0.5 by definition). 
For $T(r)$ (dashed line in Fig.1); 
$$ T(r) = 1000 ~{\rm K} 
\left( \frac{n(r)}{10^{15.0} {\rm cm^{-3}}} \right)^{\frac{1}{15}}
. \eqno(4) $$ 
According to Omukai (2000),  molecular cooling can be dominant
below a number density of $\sim 10^{15.0} {\rm cm^{-3}}$.
 According to our estimates, 
the temperature is always smaller than 1000 K. We usually set
$T_3\equiv T/1000$ K.
Line emission of H$_2$ occurs due to the changes among 
rotation and vibration states. As long as the large accretion
occurs in the range 200 K -- 1000 K (Omukai 2000),
we can safely adopt the formulation of Hollenbach \& McKee (1979)
for rotational and vibrational emission of H$_2$.
Adopting their notation: 
$$L_{\rm r} \equiv  
\left(
\frac{9.5\times 10^{-22} T_3^{3.76}}{1+0.12T_3^{2.1}} 
{\rm exp}\left[ -\left( \frac{0.13}{T_3}\right)^3 \right]
\right)
+
3.0 \times 10^{-24}{\rm exp}\left( -\frac{0.51}{T_3} \right)
~~~{\rm erg}~{\rm s}^{-1}, \eqno(5)$$
we estimate the cooling rate as 
$$\Lambda ({\rm rot}) =
n_{\rm H_2} L_{\rm r} ( 1 + \zeta_{\rm Hr} )^{-1}
+ n_{\rm H_2} L_{\rm r} ( 1 + \zeta_{\rm H_2r} ) ^{-1}
~~~{\rm erg}~{\rm cm}^{-3}~{\rm s}^{-1} ,\eqno(6)$$
where 
$\zeta_{\rm Hr} = n_{\rm Hcd}({\rm rot})/n_{\rm H}$,
$\zeta_{\rm H_2r} = n_{\rm H_2cd}({\rm rot})/n_{\rm H_2}$,
$n_{\rm Hcd}({\rm rot}) = A_J/\gamma_J^{\rm H}$, 
$n_{\rm H_2cd}({\rm rot}) = A_J/\gamma_J^{\rm H_2}$, and
$A_J$ is the Einstein $A$ value for the $J$ to $J-2$ transition;
$\gamma_J^{\rm H}$ is the collisional de-excitation rate coefficient
due to neutral hydrogen; and
$\gamma_J^{\rm H_2}$ is that due to molecular hydrogen.
The first term of $L_{\rm r}$ denotes the cooling coefficient due to 
the higher rotation level ($J>2$) and the second one due to
$J = 2 \to 0$ transition. The vibrational levels of both terms
are set to be zero.
For our parameters of interest ($250 < T < 1000$; and
$n > 10^8$ cm$^{-3}$), both  $\zeta $s are much smaller than
unity for dominant levels of $J$ because of the low temperature. 
This is because the maximum $A_J $ values are at most
$3.0 \times 10^{-7}$ sec$^{-1}$. 
As a result, these correction factors are found to be
$1 + \zeta \sim 1.0$.

For the vibrational transitions; 
$$L_{\rm v} =
6.7 \times 10^{-19} {\rm exp}\left[ -\left( \frac{5.86}{T_3}\right)^3 \right]
+
1.6 \times 10^{-18}{\rm exp}\left( -\frac{11.7}{T_3} \right)
~~{\rm erg}~{\rm s}^{-1}, \eqno(7)$$
then, we get cooling rate as being 
$$ \Lambda ({\rm vib}) = 
n_{\rm H_2} L_{\rm v} ( 1 + \zeta_{\rm Hv} )^{-1}
+ n_{\rm H_2} L_{\rm v} ( 1 + \zeta_{\rm H_2v} )^{-1}
~~{\rm erg}~{\rm cm}^{-3}~{\rm s}^{-1} ,\eqno(8)$$
where 
$\zeta_{\rm Hv} = n_{\rm Hcd}({\rm vib})/n_{\rm H}$,
and
$\zeta_{\rm H_2v} = n_{\rm H_2cd}({\rm vib})/n_{\rm H_2}$.
Here, $n_{\rm Hcd}({\rm vib}) = A_{ij}/\gamma_{ij}^{\rm H}$, and
$n_{\rm H_2cd}({\rm vib}) = A_{ij}/\gamma_{ij}^{\rm H_2}$ where
$A_{ij}$ is the Einstein $A$ value for the $i$ to $j$ transition;
$\gamma_{ij}^{\rm H}$ is collisional de-excitation rate coefficient
due to neutral hydrogen; and
$\gamma_{ij}^{\rm H_2}$ is that due to molecular hydrogen.
In our formula, only the levels of $v=0,1$ and 2 are considered.
This is sufficient since the temperature is lower than 1000 K.
The first term of $L_{\rm v}$ is a cooling coefficient of $\delta v =1$
and the second term is that of $\delta v = 2$.
The second term has no  effect on our conclusion because
of the low temperature. 
Combining Eq.(6) and Eq.(8), we obtain the total cooling rate as 
$\Lambda^{\rm thin} = \Lambda ({\rm rot}) + \Lambda ({\rm vib}) $ 
erg cm$^{-3}$ s$^{-1}$ in the optically thin regime.

We need a cooling function which can be used in the optically thick regime.
Then, adopting the following extension of $L^{\rm thin}$, 
we estimate the cooling rate as:
$$L^{\rm thick} = L^{\rm thin}
\frac{1-{\rm exp}(-\tau)}{\tau} 
{\rm exp} (-\tau_{\rm cont})
\eqno(9)$$
where $L^{\rm thin}$ is representative of 
each term of $L_{\rm r}$ and $L_{\rm v}$, 
and  $\tau$ is determined for each $L^{\rm thin}$.
The effect of the continuum absorption below 2000 K is denoted by
${\rm exp} (-\tau_{\rm cont})$, in which 
$$\tau_{\rm cont} = \rho (r) \lambda_{\rm J} 
\left[
4.1\left(\frac1{\rho (r)}-\frac1\rho_0\right)^{-0.9} {T_3}^{-4.5} 
+
0.012\rho^{0.51}(r) T_3^{2.5} 
\right]
\eqno(10)$$
according to the estimate of Lenzuni, Chernoff, \& Salpeter (1991)
who obtain a fitting formula for the Rossland mean opacity in 
a zero-metallicity gas. Here, $\lambda_{\rm J}$ is the Jeans length
and $\rho_0$ is 0.8 g cm$^{-3}$. 
Their fitting formula is reasonable
if we consider the temperature range  $T>1000$ K.
Our lowest temperature of the collapsing kernel  
is about 300 K. Then, their formula may not be appropriate, while
it gives an upper limit if we adopt $\tau_{\rm cont}$ of 1000 K
instead of really having  $\tau_{\rm cont}$ below 1000 K.
For all of our estimates, $\tau_{\rm cont}$ is sufficiently
smaller than unity. Then, we can neglect the effect of 
continuum absorption.

We shall  estimate $\tau$ 
for each  line.
Defining the suffix of $i$ as the upper energy level and
$j$ as the lower energy level, we define $\tau_{ij}$ as  $\tau$
for each  line emission.
$$ \tau_{ij} =
\frac{A_{ij}c^3}{8\pi \nu_{ij}^3}
\frac{n(x,i)g_i}{g_j}
\frac{\lambda_{\rm J}}{2\delta v_D}
.\eqno(11)$$
Here, $A_{ij}$ is a transition rate from the $i$ state to the
$j$ state;
$\nu_{ij}$ is the  central frequency of  emission in the rest frame;
$c$ is the  speed of light;
$g_i$ and $g_j$ are statistical weights for $i$ and $j$ respectively;
$\lambda_{\rm J}$ is the Jeans length
($[2k_{\rm b} T(r)/\mu(r)m_{\rm H}]^{0.5}\times [3\pi/32G\rho(r)]^{-0.5}$); 
$k_{\rm b}$ is the Boltzmann constant;
$\mu(r)$ is mean molecular weight;
and $\delta v_D = [2k_{\rm b} T(r)/\mu(r)m_{\rm H}]^{0.5}$ 
is the velocity dispersion.
The procedure of Jeans length shielding is useful for a  simple
analytical analysis (Low \& Lynden-Bell 1976; Silk 1977). 
The population density of $n(x,i)$ for rotational level
is estimated with the statistical weight of $g_J = 2J+1~(i.e.{i\to J})$. 
The suffix of $x$ denotes each  species. 
We consider $n({\rm H}_2,J)$ up to  the upper level of J=5, 
which is selected since $J=5$ is
a dominant rotational level for our fitting formulae of Eq.(1), 
Eq.(3) and Eq.(4) as checked below. 
Then, we assume $n({\rm H}_2,J>2)/n({\rm H}_2,J=2) = 27/5$

In our calculation to obtain Eq.(11), 
for a rotational transition with $v=0$,
$A_{2,0} = 2.94 \times 10^{-11}$ sec$^{-1}$; and 
$A_{J,J-2} = 5A_{2,0}/162 \times J(J-1)(2J-1)^4/(2j+1) $ 
sec$^{-1}$ are considered. For a vibrational transition,
$A_{10} = 8.3 \times 10^{-7}$ sec$^{-1}$;
$A_{21} = 1.1 \times 10^{-6}$ sec$^{-1}$; and 
$A_{20} = 4.1 \times 10^{-7}$ sec$^{-1}$ are considered.

To find the dominant rotational emission, we calculate
$$J_{\rm max} = 
\frac{\int_{\rm kernel} 4\pi r^2 n(r) J(T(r)) f_{2}(r)dr}
     {\int_{\rm kernrl} 4\pi r^2 n(r) f_{2}(r) dr}
\eqno(12)$$
where
$$
J(T(r)) = \left( \frac72 \times \frac{T(r)}{85.2~{\rm K}} \right)^{0.5}
.\eqno(13)$$
Here, $J(T(r))$ means the dominantly contributing $J$-level
to the statistical weight at temperature of $T(r)$ (Silk 1983). 
Using our fitting formula of $f_2(r)$ and $T(r)$, 
we find $J_{\rm max}$ is about 5.
Then, we regard that the dominant rotational line cooling
due  to the first term of Eq.(5) is the $J=5-3$ transition.
To estimate the luminosity
via $L^{\rm thick}$ of Eq.(9), hence, 
we use $A_{5,3}$ since the  $J=5$ ($v=0$) level is dominant. 
For  $J=2-0$ ($v=0$) transition (the second term of Eq.(5)),
we use $A_{2,0}$ to find $L^{\rm thick}$. 
The second term of Eq.(7) is always much smaller than
the other terms, and  we ignore it here. 

The estimated total luminosity 
($\int_{\rm kernel} 4\pi r^2 \Lambda dr$) with modified $L^{\rm thick}$s is
$6.7 \times 10^{34}$ erg s$^{-1}$. 
This is comparable to the luminosity estimated 
in \S2. Thus, we confirm that the large accretion rate 
is possible owing to  H$_2$ cooling. 
The emission  fraction is 0.82 for the rotational transitions with $J>2$;
0.12 for rotational transitions of $J=2-0$ ($v=0$); and 
0.06 for vibrational transition of $\delta v = 1$. 
Note that the fraction of $J>2$ is not much larger
than the fraction of $J=2-0$ ($v=0$) transition. This is caused
by  the optical depth effect. If we adopted unity for
the escape probability (i.e. any $\tau$  much smaller than unity), 
we would find the  result that there is  dominance of
$J>2$ rotational cooling. For the same reason, 
the fraction of vibrational cooling can be smaller
than that of $J=2-0$ ($v=0$). 

The line broadening is estimated to be 
$\sim \delta v_{\rm D} $
in the dimension of velocity. Adopting this,
the luminosity per Hz (T=1000 K is assumed) is the following;  
The rotation emission of $J=5-3$ 
(3.1$\times 10^{13}$ Hz; 9.7$\times 10^{-3}$ mm) is 
1.9 $\times 10^{26}$ erg Hz$^{-1}$, 
the rotation emission of $J=2-0$ 
(1.1$\times 10^{13}$ Hz; 2.8$\times 10^{-2}$ mm) is
7.9 $\times 10^{25}$ erg Hz$^{-1}$, 
and
the vibration emission of $v=1-0$ 
(1.2$\times 10^{14}$ Hz; 2.4$\times 10^{-3}$ mm) is 
3.4 $\times 10^{24}$ erg Hz$^{-1}$.

\section{OBSERVATIONAL FEASIBILITY}

In the current paper, we need the observational frequency and
wavelength which are redshifted. For the three lines;
we get
1.6$\times 10^{12}$ Hz; 1.9$\times 10^{-1}$ mm for $J=5-3$,
0.53$\times 10^{12}$ Hz; 5.6$\times 10^{-1}$ mm for $J=2-0$,
and
6.1$\times 10^{12}$ Hz; 4.9$\times 10^{-2}$ mm for $\delta v = 1$.
The adopted $1+z$ is 20 according to the results of
recent numerical simulations.

To find the observed flux, we determine the distance to
the object. Then, we calculate it numerically
from the following standard formula: 
$$ D_{19} = \frac{c}{H_0}
\int_0^{19} \frac{dz}{(\Omega_\Lambda + 
                       \Omega_{\rm M}(1+z)^{3.0})^{0.5}}
\eqno(14)$$
where  $H_0$ is the Hubble parameter and used 75 km sec$^{-1}$ Mpc$^{-1}$, 
$\Omega_\Lambda$ is the cosmological constant parameter, and
$\Omega_{\rm M}$ is the density parameter.
We consider the case  $\Omega_\Lambda + \Omega_{\rm M} =1$.
Adopting $D_{19}$, we obtain the observed fluxes of each of the lines.
The results are  summarized in table 1.
Although the line-broadening is estimated from $\nu_0 \delta v_{\rm D} /c$,
we also correct it with redshift effect 
($\nu_0$ is the  central  frequency).
For each of the parameter sets of ($\Omega_\Lambda,\Omega_{\rm M}$),
we obtain $D_{19} = 0.62 \times 10^{10}$ pc 
($\Omega_\Lambda=0,\Omega_{\rm M}=1$),
$D_{19} = 1.00\times 10^{10}$ pc
($\Omega_\Lambda=0.7,\Omega_{\rm M}=0.3$),
and 
$D_{19} = 4.58 \times 10^{10}$ pc
($\Omega_\Lambda=0.9,\Omega_{\rm M}=0.1$).
According to Table 1, the rotational line of $J=5-4$ (v=0)
is 0.07 $\mu$Jy; 0.05 $\mu$Jy for $J=2-0$ (v=0); 
and 0.001 $\mu$Jy for $\delta v = 1$ 
if $\Omega_{\rm M} =1$. 
Thus, we find that a single kernel is not observable
by any available observational facilities in the near future.
The expected flux is too small.

However we note  that if the kernels are  assembled on a  galactic scale,
the assembly can be detected by  ongoing projects.
We shall estimate the  number of kernels in a primordial galaxy.
Firstly, we must consider the lifetime of a H$_2$ emission kernel.
It may be $10^4$ years for a 100 $M_\odot$ star at the
specified accretion rate
of 0.01$M_\odot$ year$^{-1}$.
A $10^6 M_\odot$ cloud is expected 
to  form 1000 such stars at  a
plausible efficiency.  Taking its lifetime to be $10^5$ years 
as the  free-fall time of  a $10^6 M_\odot$ cloud,
we find that its luminosity is 100 $L_{\rm acc}$ 
(see Eq.(2) for definition of $L_{\rm acc}$).

Next, we consider an entire primordial galaxy  with $10^{11} M_\odot$.
It may form $10^9$ supernovae over its entire lifetime.
We assume that it makes $10^7$ primordial massive stars, since 
this  number of massive stars gives enrichment to  roughly
1 percent of 
solar metallicity. 
If the burst of formation of primordial molecular cloud kernels
occurs in the central region of the primordial galaxy (we postulate  
1 kpc as the size of the  kernel-forming region), then 
there are $10^4$ such giant molecular clouds with $10^6 M_\odot$ 
in the kernel-forming region. During  this phase,
the cumulative luminosity would be $10^6$ $L_{\rm acc}$.  
It seems that the predicted line luminosity would be  easy to  observe.
However, the dynamical time-scale in the kernel forming region
may be $10^7$ years (e.g. the duration of the starburst).
Then, we obtain $10^4~L_{\rm acc}$ for the luminosity of 
the primordial galaxy undergoing its first star formation burst,
conservatively assuming $10^5$ years as the life-time 
of giant molecular clouds.
Finally, for H$_2$ emitting proto-galaxies,
we obtain  0.7 mJy for $J=5-3$, 0.5 mJy for $J=2-0$ 
and 0.01 mJy for $\delta v =1$ ($\Omega_{\rm M} = 1.0$).

Thus, the rotational emissions of $J=5-3$ and $J=2-0$
are at sub-mJy flux levels. Those frequencies enter into the
sub-mm and far-infrared
bands if the kernels form at $1+z=20$. These are observational
ranges of ALMA and ASTRO-F. 
According to the current status of the instrumentation
for ALMA,  80-890 GHz is the allowed detectable range.
It is feasible for our prediction of the redshifted H$_2$ emission
of the $J=2-0$ line. Unfortunately, however, the transmission is bad
for the predicted feature at 530 GHz because of the atmosphere of the Earth. 
Then, it may be necessary to detect  H$_2$ emission from 
a  protogalaxy forming later than $1+z=20$. 
We note that `$1+z=20$' is obtained by taking $\Omega_M = 1$. 
If $\Omega_M < 1$, which seems to be reasonable if $\Omega_\Lambda = 0.7$,
the formation of kernels is delayed. Then, we can expect 
to detect emission at a  higher frequency than 530 GHz and 
detection should be feasible of  the H$_2$ emmision from the assembly of 
the primordial molecular kernels.

For the case of the $J=5-3$ line, this is observable at 194 $\mu$m wave-length.
This is  in the allowed range of ASTRO-F (2-200 $\mu$m). 
At the wavelength of 200 $\mu$m, unfortunately, it is
difficult to detect  sub mJy-level fluxes because of the properties
of the detectors. 
If a higher rotational transition than $J=5-3$ dominates,
the observations are  more suitable for the allowed range of ASTRO-F.
The higher transitions emit more energetic photons,
and  enter into the possible range of ASTRO-F.
This may be possible if the external radiation field
(e.g. Susa \& Umemura 2000) has positive feedback
on  the primordial kernels. 
Thus, although we need a very specific condition in order to detect
H$_2$ emitting protogalaxies, it may be possible
to achieve this goal  with  future observational facilities. 

Here, we advocate a deep blank field survey.  The unit area for the
survey can be estimated by the number density of QSOs, since QSOs are
 believed to form at high density peaks during  cosmic
structure formation.  The burst of  primordial kernel formation may
be expected in the environment where QSOs can form. The most recent
and reliable number density of QSOs is determined by Miyaji, Hasinger
\& Schmidt (2000). According to their luminosity function of soft
X-ray AGNs, a typical number density (e.g. their figure 11) is about
$10^{-6}$ $h_{50}^3$ Mpc$^{-3}$ beyond $z\sim 3$. They have assumed that
Hubble parameter is 50 km/s/Mpc.  Then, a unit covering area for a single QSO
at a fixed redshift (i.e. a fixed distance) 
is typically $10^{4}$ $h_{50}^{-2}$ Mpc$^{2}$. The
proper transverse size of an area, $D_{\rm size}$, as seen by us, is its
comoving size times the scale correction factor at the time of emission;
$\delta \theta D_{19} / 20$ at $1+z=20$, where $\delta \theta$ has the
units of radians.  Hence the unit area needed for our blank field
survey is
estimated to be $7.6 \times 7.6$ degree$^2$ if $D_{\rm size} = 66.6$ Mpc,
$D_{19} = 10^{10}$ pc, and $H_0 = 75$.  The field of view of ALMA is
about $10 \times 10$ arcsec$^2$.  Thus, we need a
huge amount of observing  time to
survey the expected area.  The field  of ASTRO-F for the long wavelength
detector is $12.5 \times 2.5$ arcmin$^2$.  Although ASTRO-F plans to
survey nearly all the sky, the detection limit for the survey mode is at
10 mJy level and we cannot expect to detect H$_2$ emission.  We need good
pointing and long exposure-time for  the entire area of $7.6 \times 7.6$
degree$^2$. Then, the pointing mode of ASTRO-F may be suitable.  It
can detect line emission at the mJy level in a single
 pointing exposure.  To
detect the predicted sub-mJy level flux, we need to expose the same region several times.  As a result, we can expect  that an $H_2$
emitting protogalaxy could be  found near the oldest QSO via   deep
and off-centred (i.e. avoiding saturation) spectroscopy using  the pointing
mode of ASTRO-F. 

It might be difficult to justify the necessary observing time
for a blank field survey.
In such a situation, we  need to specify the observational target.
Which object is suitable for our first observations? 
One possibility would be unidentified SCUBA sources 
(e.g. Smail, Ivison, \& Blain 1997). These are believed to  have been
detected at high redshift owing to the so-called
"negative k-correction". As long as they are primordial galaxies,
the possible range of $z$ for SCUBA objects 
is from 3 to 10. If $z=10$-SCUBA objects
are observed by near-future facilities,
it may be possible to detect emission from   
H$_2$ at the sub-mJy level. But, in this case, the background radiation
from dust prevents us from confirming H$_2$ line
emission. To resolve this, a
high dispersion spectrograph is needed.
Furthermore, if the SCUBA objects are  metal-rich primordial galaxies,
the primordial molecular cloud evolution favours the fragmentation 
scenario (Bromm et al. 2001), and low mass star formation.  
The criterion  of Bromm et al.
for  being at metallicity below  $10^{-3.5} \times Z_\odot$
to allow massive primordial star formation
is only marginally satisfied at this redshift, as would be required  
to allow   dominant H$_2$ cooling as suggested by Omukai (2000). 

\section{SUMMARY}

One of the main goals of cosmology is  to find the first generation
of stars. When they form,
strong H$_2$ emission is expected. We have examined the
observational feasibility of detection. 
According to our analysis, $J=5-3$ and $J=2-0$ rotational
emission are expected (at $v=0$).
Protogalaxies can have sub-mJy fluxes when the
first stars form at $1+z=20$
as an assembly.
It seems that these systems  are marginally detectable
by ALMA and ASTRO-F.
However, the transmissivity of air for the emission of $J=2-0$
is low for the ALMA project, and 
the wavelength edge of the filter for the far-infrared band is not 
sufficiently sensitive  for the $J=5-3$ line.
Thus if  future telescopes can detect H$_2$ emission 
from   primordial molecular cloud kernels,
the kernels and first stars have formed after $1+z=20$ 
and/or the temperature of the kernel should be larger
than predicted in the simplest models.

\acknowledgments
H.K. is grateful to Profs. S.Mineshige, S.Inagaki and R.Hirata
for their encouragement.

%
%

\clearpage

\begin{table}
\begin{center}
\caption{Expected Emission Lines} 
\begin{tabular}{crrrccc}
\tableline\tableline
--- & $J=5-3$ & $J=2-0$ & $\delta v = 1$ 
    & $\Omega_m$ & $\Omega_\Lambda$  & $D_{19}$ ($10^{10}$ pc)\\
\tableline
$\nu$     (10$^{13}$ Hz; $z=0$) &3.09 &1.06 &12.21 & -- & -- & --\\
$\nu$     (10$^{12}$ Hz; $z=19$)&1.55 &0.53 &6.11  & -- & -- & --\\
$\lambda$ (10$^{-3}$ mm; $z=0$) &9.71 &28.29 &2.46 & -- & -- & --\\
$\lambda$ (10$^{-3}$ mm; $z=19$)&194.2 &565.8 &49.2& -- & -- & --\\
$L_\nu$   (10$^{26}$ erg sec$^{-1}$ Hz$^{-1}$) 
                       &1.87 &0.79  &0.034 &-- & -- & --  \\
$f_\nu$   (10$^{-8}$ Jy; $z=19$) 
                        &7.34 &5.32 & 0.148& 1.0 & 0.0 & 0.62 \\ 
$f_\nu$   (10$^{-8}$ Jy; $z=19$)
                        &2.85 &2.07 & 0.058& 0.3 & 0.7 & 1.00 \\
$f_\nu$   (10$^{-8}$ Jy; $z=19$)
                        &1.29 &0.94 & 0.026& 0.1 & 0.9 & 1.48 \\
\tableline
\end{tabular}
\end{center}
\end{table}

\clearpage

\centerline{FIGURE CAPTIONS}

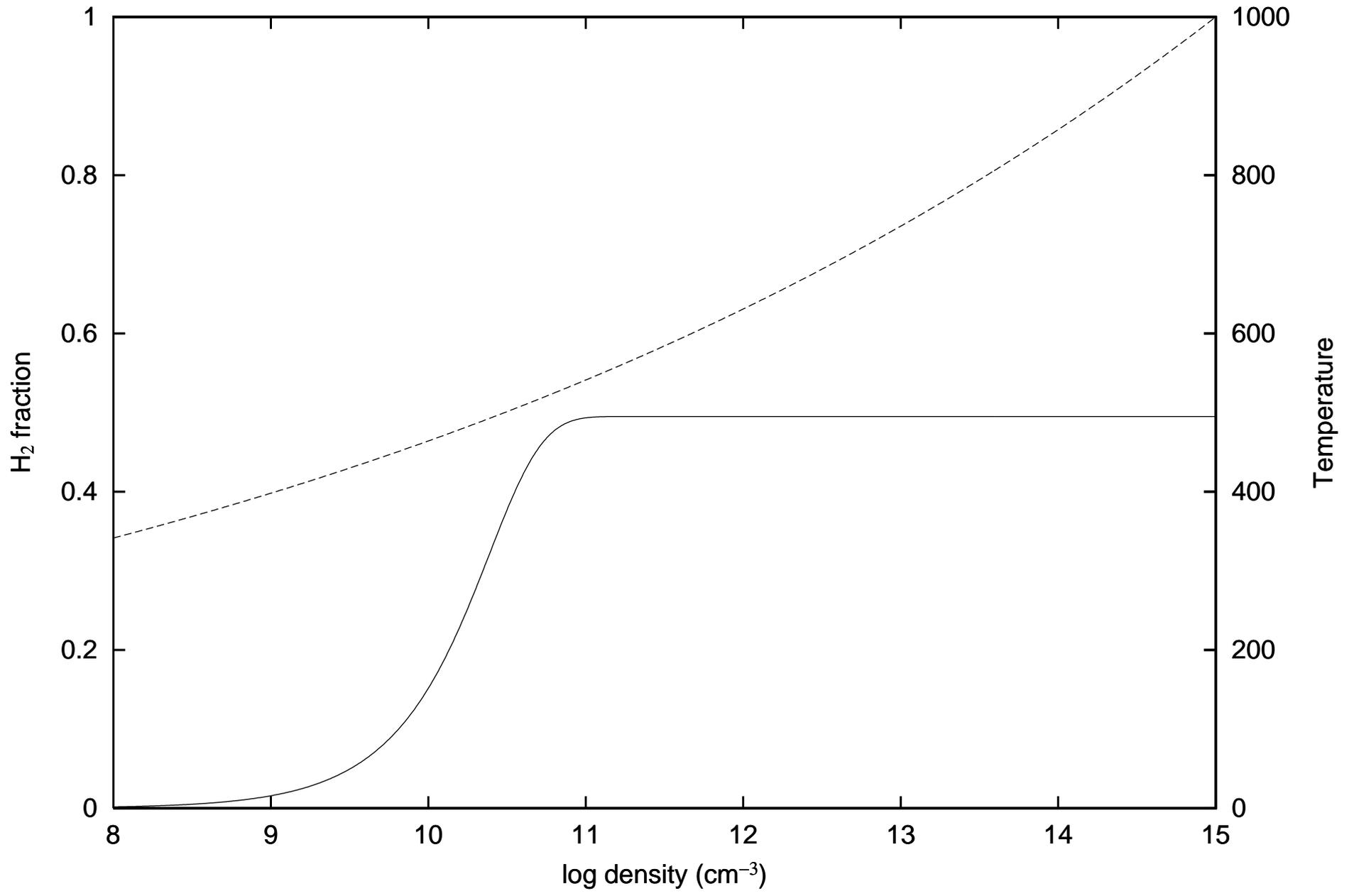
Fig.1 ---  Density--H$_2$ fraction relation (solid line) and
Density--Temperature relation (dashed line) are depicted.
Using Eq.(1),
we translate them to Radius--Density and --Temperature relations,
respectively.

\end{document}